\documentstyle[12pt,epsf]{article}
\newcommand{\nc}{\newcommand}
\nc{\postscript}[2] 
{\setlength{\epsfxsize}{#2\hsize}\centerline{\epsfbox{#1}}}
\nc{\bg}{B. Grz\c{a}dkowski}
\nc{\wj}{P_{e^-}} \nc{\wdw}{P_{e^+}} \nc{\sktw}{\sin^2\theta_W}
\nc{\cktw}{\cos^2\theta_W}     \nc{\sd}{|S^D\,|^2}
\nc{\sdrl}{|S^D_{RL}\,|^2}     \nc{\td}{|\,T^D\,|^2}
\nc{\vd}{|\,V^D\,|^2}          \nc{\st}{\,{\rm Re}(S^D T^{D*})}
\nc{\s}{|S\,|^2}               \nc{\tp}{|\,T\,|^2}
\nc{\fjed}{\frac{s^2}{128\pi^2 \alpha^2}}
\nc{\fdwa}{\frac{4\pi g^4W}{m_t^2M_W{\mit\Gamma}_W}}
\nc{\dvp}{D'_V}   \nc{\dap}{D'_A}    \nc{\dvap}{\,{\rm Re}D'_{V\!A}}
\nc{\mianalfa}{G+2(\sd +\sdrl +4\vd +12\td)}
\nc{\wvt}{v_t}    \nc{\wet}{e_t}     \nc{\wae}{a_e}
\nc{\wat}{a_t}    \nc{\al}{A_L}      \nc{\ar}{A_R}
\nc{\bl}{B_L}     \nc{\br}{B_R}      \nc{\wsp}{\frac{s}{e^2}}
\nc{\fvv}{\wve\wvt d-\wet}           \nc{\fav}{\wae\wvt d}
\nc{\fva}{\wve\wat d}                \nc{\faa}{\wae\wat d}
\nc{\fvvx}{\fvv+\frac{A_L+A_R}{4}\wsp}
\nc{\favx}{\fav+\frac{A_L-A_R}{4}\wsp}
\nc{\fvax}{\fva+\frac{B_L+B_R}{4}\wsp}
\nc{\faax}{\faa+\frac{B_L-B_R}{4}\wsp}
\nc{\jww}{(1+{\wj}{\wdw})}           \nc{\ww}{({\wj}+{\wdw})}
\nc{\fdv}{\Bigl|\fvvx\Bigr|^2+\Bigl|\favx\Bigr|^2}
\nc{\fda}{\Bigl|\fvax\Bigr|^2+\Bigl|\faax\Bigr|^2}
\nc{\fdva}{(\fvax)(\fvvx)^*\\&&\lspace\lspace+(\faax)(\favx)^*}
\nc{\fev}{\:{\rm Re}\Bigl[(\fvvx)(\favx)^*\Bigr]}
\nc{\fea}{\:{\rm Re}\Bigl[(\fvax)(\faax)^*\Bigr]}
\nc{\feva}{(\fvax)(\favx)^*\\&&\lspace\lspace+(\faax)(\fvvx)^*}
\nc{\rest}{{\rm Re}(ST^*)}  \nc{\imst}{{\rm Im}(ST^*)}   
\nc{\invlq}{lq}             \nc{\invP}{Ps_++Ps_-}
\nc{\invl}{ls_+-ls_-}       \nc{\invppll}{Ps_+\;Ps_--ls_+\;ls_-}
\nc{\invlppl}{ls_+\;Ps_--Ps_+\;ls_-}
\nc{\invlpplm}{ls_-\;Ps_+-ls_+\;Ps_-}
\nc{\invpllp}{Ps_-\;ls_+-ls_-\;Ps_+}
\nc{\invllpp}{ls_+\;ls_--Ps_+\;Ps_-}
\nc{\invpm}{Ps_--Ps_+}      \nc{\invpp}{Ps_+-Ps_-}
\nc{\invlp}{ls_-+ls_+}      \nc{\invlm}{-ls_--ls_+}
\nc{\ssdot}{s_+s_-}         \nc{\jmbk}{1-\beta^2}
\nc{\njp}{N_{1+}}           \nc{\njm}{N_{1-}}
\nc{\ndp}{N_{2+}}           \nc{\ndm}{N_{2-}}
\nc{\jd}{\frac{1}{2}}       \nc{\dvk}{\,D^{(*)}_V}
\nc{\dak}{\,D^{(*)}_A}      \nc{\dvak}{\,D^{(*)}_{V\!A}}
\nc{\redvak}{\,{\rm Re}(\dvak)}
\nc{\imdvak}{\,{\rm Im}(\dvak)}
\nc{\evk}{\,E^{(*)}_V}      \nc{\eak}{\,E^{(*)}_A}
\nc{\evak}{\,E^{(*)}_{V\!A}}
\nc{\reevak}{\,{\rm Re}(\evak)}
\nc{\imevak}{\,{\rm Im}(\evak)}
\nc{\mianeta}{(3-\beta^2)\dvk +2\beta^2\dak
    +\zeta_1[\,3(1+\beta^2)\s +4(3-\beta^2)\tp\,]} 
\nc{\wve}{v_e}              \nc{\jnl}{\frac{1}{{\mit\Lambda}^2}}
\nc{\jc}{\frac{1}{4}}  \nc{\sll}{S_{LL}}     \nc{\slr}{S_{LR}}
\nc{\srl}{S_{RL}}      \nc{\srr}{S_{RR}}     \nc{\vll}{V_{LL}}
\nc{\vlr}{V_{LR}}      \nc{\vrl}{V_{RL}}     \nc{\vrr}{V_{RR}}
\nc{\tll}{T_{LL}}      \nc{\tlrs}{T_{LR}}    \nc{\trl}{T_{RL}}
\nc{\trr}{T_{RR}}      \nc{\slld}{S_{LL}^D}  \nc{\slrd}{S_{LR}^D}
\nc{\srld}{S_{RL}^D}   \nc{\srrd}{S_{RR}^D}  \nc{\vlld}{V_{LL}^D}
\nc{\vlrd}{V_{LR}^D}   \nc{\vrld}{V_{RL}^D}  \nc{\vrrd}{V_{RR}^D}
\nc{\tlld}{T_{LL}^D}   \nc{\tlrd}{T_{LR}^D}  \nc{\trld}{T_{RL}^D}
\nc{\trrd}{T_{RR}^D}   \nc{\aqde}{\alpha_{qde}}
\nc{\alq}{\alpha_{\ell q}}        \nc{\alqp}{\alpha_{\ell q'}}
\nc{\alqt}{\alpha_{\ell q}^{(3)}} \nc{\alqtc}{\alpha_{\ell q}^{(3)*}}
\nc{\alqj}{\alpha_{\ell q}^{(1)}} \nc{\alqjc}{\alpha_{\ell q}^{(1)*}}
\nc{\aeu}{\alpha_{eu}}      \nc{\alu}{\alpha_{\ell u}}
\nc{\aqe}{\alpha_{qe}}      \nc{\ber}{\begin{eqnarray*}}
\nc{\enr}{\end{eqnarray*}}  \nc{\jmpb}{(1-\beta)/(1+\beta)}
\nc{\wspR}{r}      \nc{\varx}{x}      \nc{\bt}{\beta}
\nc{\tpw}{\frac{3}{W}}      \nc{\tom}{\;\theta(1-r-\omega)}
\nc{\tx}{\;\theta\Bigl(x-r\frac{1-\beta}{1+\beta}\Bigr)}
\nc{\txp}{\;\theta\Bigl(x'-r\frac{1-\beta}{1+\beta}\Bigr)}
\nc{\non}{\nonumber}
\nc{\barx}{\bar{x}}         \nc{\pbarn}{\;\hbox {pb}}
\nc{\fbarn}{\;\hbox {fb}}   \nc{\hc}{\hbox {h.c.}}
\nc{\re}{\hbox {Re}}        
\nc{\mev}{\hbox {MeV}} \nc{\gev}{\;\hbox {GeV}}
\nc{\tev}{\;\hbox {TeV}}
\def\gesim{\lower0.5ex\hbox{$\:\buildrel >\over\sim\:$}} 
\def\lesim{\lower0.5ex\hbox{$\:\buildrel <\over\sim\:$}}
\nc{\app}[3]{{\it Acta\ Phys.\ Pol.}\ {{\bf B{#1}} (#2), #3}}
\nc{\prd}[3]{{\it Phys.\ Rev.}\ {{\bf D{#1}} (#2), #3}}
\nc{\prl}[3]{{\it Phys.\ Rev.\ Lett.}\ {{\bf {#1}} (#2), #3}}
\nc{\plb}[3]{{\it Phys.\ Lett.}\ {{\bf B{#1}} (#2), #3}}
\nc{\npb}[3]{{\it Nucl.\ Phys.}\ {{\bf B{#1}} (#2), #3}}
\nc{\ptp}[3]{{\it Prog.\ Theor.\ Phys.}\ {{\bf {#1}} (#2), #3}}
\nc{\zfp}[3]{{\it Z.\ Phys.}\ {{\bf C{#1}} (#2), #3}}
\nc{\mpla}[3]{{\it Mod.\ Phys.\ Lett.}\ {{\bf A{#1}} (#2), #3}}
\nc{\rmp}[3]{{\it Rev.\ Mod.\ Phys.}\ {{\bf {#1}} (#2), #3}}
\nc{\ijmpa}[3]{{\it Int.\ J.\ of\ Mod.\ Phys.}\
               {{\bf A{#1}} (#2), #3}}
\nc{\ttbar}{t\bar{t}}         \nc{\bbbar}{b\bar{b}}
\nc{\tanb}{\tan \beta}        \nc{\twbdec}{t\to W^+ b}
\nc{\tbwbdec}{\bar{t}\to W^- \bar{b}}
\nc{\epem}{e^+e^-}            \nc{\eett}{\epem \to \ttbar}
\nc{\sigeett}{\sigma_{e\bar{e}\to\ttbar}}
\nc{\wpwm}{W^+W^-}            \nc{\tbar}{\bar{t}}
\nc{\bbar}{\bar{b}}           \nc{\wpp}{W^+}
\nc{\mt}{m_t}    \nc{\mts}{m_t^2}   \nc{\mw}{M_W}    \nc{\mws}{M_W^2}
\nc{\mz}{M_Z}    \nc{\mzs}{M_Z^2}
\nc{\ttbardec}{\ttbar \to W^+W^-\bbbar}
\nc{\wwbb}{W^+W^-\bbbar}      \nc{\sm}{S\!M}
\nc{\cw}{\cos\theta_W}        \nc{\sw}{\sin\theta_W}
\nc{\sws}{\sin^2\theta_W}     \nc{\sig}{\sigma_{tot}}
\nc{\lp}{\ell^+}              \nc{\lm}{\ell^-}
\nc{\epsl}{\epsilon_L}        \nc{\cp}{C\!P}

\nc{\splus}{s_+}       \nc{\smin}{s_-}        \nc{\eps}{\epsilon}
\nc{\psp}{Ps_+}        \nc{\psm}{Ps_-}        \nc{\lsp}{ls_+}
\nc{\lsm}{ls_-}        \nc{\sss}{s_+s_-}      \nc{\m}{m_t}
\nc{\mq}{m_t^2}        \nc{\mr}{\frac{1}{\m}} \nc{\av}{A_{\gamma}}
\nc{\bv}{B_{\gamma}}   \nc{\az}{A_Z}          \nc{\bz}{B_Z}
\nc{\avs}{A_{\gamma}^2}\nc{\azs}{A_Z^2}       \nc{\bzs}{B_Z^2}
\nc{\dav}{\delta \! A_{\gamma}}   \nc{\dbv}{\delta \! B_{\gamma}}
\nc{\dcv}{\delta C_{\gamma}}      \nc{\ddv}{\delta \! D_{\gamma}}
\nc{\daz}{\delta \! A_Z}          \nc{\dbz}{\delta \! B_Z}
\nc{\dcz}{\delta C_Z}             \nc{\ddz}{\delta \! D_Z}
\nc{\dev}{\delta \! E_{\gamma}}   \nc{\dez}{\delta \! E_Z}
\nc{\dfv}{\delta \! F_{\gamma}}   \nc{\dfz}{\delta \! F_Z}
\nc{\rdav}{{\rm Re}(\delta \! A_{\gamma}) \:}
\nc{\rdbv}{{\rm Re}(\delta \! B_{\gamma}) \:}
\nc{\rdcv}{{\rm Re}(\delta C_{\gamma}) \:}
\nc{\rddv}{{\rm Re}(\delta \! D_{\gamma}) \:}
\nc{\rdaz}{{\rm Re}(\delta \! A_Z) \:}
\nc{\rdbz}{{\rm Re}(\delta \! B_Z) \:}
\nc{\rdcz}{{\rm Re}(\delta C_Z) \:}
\nc{\rddz}{{\rm Re}(\delta \! D_Z) \:}
\nc{\idav}{{\rm Im}(\delta \! A_{\gamma}) \:}
\nc{\idbv}{{\rm Im}(\delta \! B_{\gamma}) \:}
\nc{\idcv}{{\rm Im}(\delta C_{\gamma}) \:}
\nc{\iddv}{{\rm Im}(\delta \! D_{\gamma}) \:}
\nc{\idaz}{{\rm Im}(\delta \! A_Z) \:}
\nc{\idbz}{{\rm Im}(\delta \! B_Z) \:}
\nc{\idcz}{{\rm Im}(\delta C_Z) \:}
\nc{\iddz}{{\rm Im}(\delta \! D_Z) \:}
\nc{\cz}{(1+v_e^2)d\:\!'^2}         \nc{\ci}{v_ed\:\!'}
\nc{\ccz}{v_ed\:\!'^2}              \nc{\cci}{d\:\!'}
\nc{\lspace}{\;\;\;\;\;\;\;\;\;\;}  \nc{\llspace}{\lspace \lspace}

\nc{\beq}{\begin{equation}}   \nc{\eeq}{\end{equation}}
\nc{\bea}{\begin{eqnarray}}   \nc{\eea}{\end{eqnarray}}
\nc{\baa}{\begin{array}}      \nc{\eaa}{\end{array}}
\nc{\bit}{\begin{itemize}}    \nc{\eit}{\end{itemize}}
\nc{\ben}{\begin{enumerate}}  \nc{\een}{\end{enumerate}}
\nc{\bce}{\begin{center}}     \nc{\ece}{\end{center}}
\begin{document}
\pagestyle{empty} \setlength{\footskip}{2.0cm}
\setlength{\oddsidemargin}{0.5cm} \setlength{\evensidemargin}{0.5cm}
\renewcommand{\thepage}{-- \arabic{page} --}
\def\mib#1{\mbox{\boldmath $#1$}}
\def\bra#1{\langle #1 |}      \def\ket#1{|#1\rangle}
\def\vev#1{\langle #1\rangle} \def\dps{\displaystyle}
   \def\thebibliography#1{\centerline{REFERENCES}
     \list{[\arabic{enumi}]}{\settowidth\labelwidth{[#1]}\leftmargin
     \labelwidth\advance\leftmargin\labelsep\usecounter{enumi}}
     \def\newblock{\hskip .11em plus .33em minus -.07em}\sloppy
     \clubpenalty4000\widowpenalty4000\sfcode`\.=1000\relax}\let
     \endthebibliography=\endlist
   \def\sec#1{\addtocounter{section}{1}\section*{\hspace*{-0.72cm}
     \normalsize\bf\arabic{section}.$\;$#1}\vspace*{-0.3cm}}
\vspace*{-1cm}\noindent
\hspace*{10.8cm}IFT-16-97\\
\hspace*{10.8cm}TOKUSHIMA 97-02\\
\hspace*{10.8cm}(hep-ph/9712357)\\

\vspace*{.5cm}

\begin{center}
{\large\bf Four-Fermi Effective Operators in Top-Quark}

\vskip 0.16cm
{\large\bf Production and Decay}
\end{center}

\vspace*{0.8cm}
\begin{center}
\renewcommand{\thefootnote}{\alph{footnote})}
{\sc Bohdan GRZ\c{A}DKOWSKI$^{\:1),\:}$}\footnote{E-mail address:
\tt bohdan.grzadkowski@fuw.edu.pl},\
{\sc Zenr\=o HIOKI$^{\:2),\:}$}\footnote{E-mail address:
\tt hioki@ias.tokushima-u.ac.jp}, 

\vskip 0.15cm
{\sc Micha{\l} SZAFRA{\'N}SKI$^{\:1),\:}$}\footnote{E-mail
address: \tt michal.szafranski@fuw.edu.pl}
\end{center}

\vspace*{1.2cm}
\centerline{\sl $1)$ Institute for Theoretical Physics,\ Warsaw 
University}
\centerline{\sl Ho\.za 69, PL-00-681 Warsaw, POLAND} 

\vskip 0.3cm
\centerline{\sl $2)$ Institute of Theoretical Physics,\ 
University of Tokushima}
\centerline{\sl Tokushima 770-8502, JAPAN}

\vspace*{2.2cm}
\centerline{ABSTRACT}

\vspace*{0.4cm}
\baselineskip=20pt plus 0.1pt minus 0.1pt
Effects of four-Fermi-type new interactions are studied in top-quark
pair production and their subsequent decays at future $e^+e^-$
colliders. Secondary-lepton-energy distributions are calculated for
arbitrary longitudinal beam polarizations. An optimal-observables
procedure is applied for the determination of new parameters.
\vfill
\newpage
\renewcommand{\thefootnote}{\sharp\arabic{footnote}}
\pagestyle{plain} \setcounter{footnote}{0}
\baselineskip=21.0pt plus 0.2pt minus 0.1pt

\sec{Introduction}

The Standard Model of the electroweak interactions (SM) has so far
never failed in describing various low- and high-energy phenomena in
particle physics. In spite of this success, however, a more
fundamental theory is desired in order to eliminate arbitrariness
embedded in the SM. Once we assume a specific model, e.g. a SUSY
model as a candidate, we will be able to calculate cross sections
and/or decay widths and test the model comparing predictions with
experimental data. Here, however, we will follow a general
model-independent strategy adopting an effective lagrangian~\cite{bw}
to describe non-standard physics. We will discuss thereby an
influence of beyond-the-SM interactions on a production and decay of
top quarks at future $e^+e^-$ colliders (NLC).

In our approach non-standard interactions are parameterized in terms
of a set of effective local operators that respect symmetries of the
SM. The operators are gauge invariant with canonical dimension $>$ 4.
In order to write down the effective lagrangian explicitly, we have
to choose the low-energy particle content. Here we will assume that
the SM spectrum correctly describes all such excitations. Thus we
imagine that there is a scale ${\mit\Lambda}$, at which new physics
becomes apparent, and all new effects are suppressed by inverse
powers of ${\mit\Lambda}$. A catalogue of the operators up to
dimension 6 is given in \cite{bw}.

Some of the new interactions in the effective lagrangian generate
corrections to the SM couplings like $\gamma q\bar{q}$, $Zq\bar{q}$,
$Wqq'$ etc.. In our recent works \cite{GH_npb,GH_plb,BGH}, we have
discussed consequences of modified vector-boson couplings to
fermions. In this paper, we shall focus on four-Fermi interactions
and study their effects on the secondary-lepton-energy distributions
in the process $\epem \to \ttbar \to \ell^\pm \cdots$. In section~2,
we list all four-Fermi operators and present the corresponding
effective lagrangian which contribute to $e^+e^-\to t\bar{t}$ and
$t\to b\ell^+\nu_{\ell}/\bar{t}\to\bar{b}\ell^-\bar{\nu}_{\ell}$. In
section 3 we derive the secondary-lepton-energy distributions, and in
section 4 we apply the optimal observable procedure \cite{optobs} to
determine couplings of the four-Fermi operators. We summarize our
results in the final section. In the appendix we present explicit
formulas for the angular distribution of polarized top quarks
produced at $e^+e^-$ scattering (A), the decay width of $t$ and
$\bar{t}$ (B) and some relevant functions used for the energy
spectrum of secondary leptons (C and D).

\sec{Four-Fermi effective operators}

\noindent
{\bf a. $\mib{t}\bar{\mib{t}}$ production} \\
Let us start with $e^+e^-\to t\bar{t}$. The following
tree-level-generated operators \cite{gone} will directly contribute
to this process:
\begin{equation}
\begin{array}{lcllcl}
{\cal O}^{(1)}_{\ell q}&\!\!=&\!\!\dps\frac12
(\bar{\ell}\gamma_{\mu}\ell)(\bar{q}\gamma^{\mu}q),\;\;\;&
{\cal O}^{(3)}_{\ell q}&\!\!=&\!\!\dps\frac12
(\bar{\ell}\gamma_{\mu}\tau^I\ell)
(\bar{q}\gamma^{\mu}\tau^Iq),\\ \vspace*{-0.3cm}
& & & & & \\ \vspace*{-0.3cm}
{\cal O}_{eu}&\!\!
=&\!\!\dps\frac12
(\bar{e}\gamma_{\mu}e)(\bar{u}\gamma_{\mu}u), & & & \\
& & & & & \\
{\cal O}_{\ell u}&\!\!=&\!\!(\bar{\ell}u)(\bar{u}\ell),&  
{\cal O}_{qe}&\!\!=&\!\!(\bar{q}e)(\bar{e}q),\\
{\cal O}_{\ell q}&\!\!=&\!\!(\bar{\ell}e)\epsilon (\bar{q}u), &
{\cal O}_{\ell q'}&\!\!=&\!\!(\bar{\ell}u)\epsilon (\bar{q}e).
\end{array}
\end{equation}
Given the above list the lagrangian which we will use in the
following calculations is:
\begin{equation}
{\cal L}={\cal L}^{S\!M}+
\frac{1}{{\mit\Lambda}^2}\sum_i(\:\alpha_i{\cal O}_i+{\rm h.c.}\:),
\end{equation}
where $\alpha$'s are the coefficients which parameterize non-standard
interactions. It should be emphasized that, according to the
classification developed in ref.~\cite{wudka}, coefficients in front
of four-Fermi operators may be large since the operators could be
generated at the tree level of perturbation expansion within certain
underlying theory.\footnote{Assuming the underlying theory is a
gauge theory and the perturbative expansion is justified.}

After Fierz transformation the part of lagrangian containing the
above four-Fermi operators can be rewritten as follows \cite{gone}:
\begin{eqnarray}
&&{\cal L}^{4F}
 =\sum_{i,j=L,R}\Bigl[\:S_{ij}(\bar{e}P_ie)(\bar{t}P_jt) \non \\
&&\lspace
 +V_{ij}(\bar{e}\gamma_{\mu}P_ie)(\bar{t}\gamma^{\mu}P_jt)+T_{ij}
 (\bar{e}\frac{\sigma_{\mu\nu}}{\sqrt{2}}P_ie)
(\bar{t}\frac{\sigma^{\mu\nu}}{\sqrt{2}}P_jt)\:\Bigr] \label{dec_lag}
\end{eqnarray}
with the following constraints satisfied by the coefficients:
\begin{eqnarray*}
&&S_{RR}=S^{*}_{LL},\ \ \ S_{LR}=S_{RL}=0,\\
&&\ \ \ \ \ \ \ \ \ \ \ \ V_{ij}=V^{*}_{ij},\\
&&T_{RR}=T^{*}_{LL},\ \ \ T_{LR}=T_{RL}=0,
\end{eqnarray*}
where 
\renewcommand{\arraystretch}{1.9}
\begin{equation}
\begin{array}{lcllcl} \;\;\;\;\;
\sll&\!\!=&\!\!\dps\jnl(-\alq^{*}+\jd\alqp^{*}),& \!\!\!\!\!\!\!\!\!
\srr&\!\!=&\!\!\dps\jnl(-\alq+\jd\alqp), \\ \;\;\;\;\;
\vll&\!\!=&\!\!\dps\jd\dps\jnl(\alqj-\alqt+\alqjc-\alqtc),& & &
\\ \;\;\;\;\;
\vlr&\!\!=&\!\!-\dps\jd\dps\jnl(\alu+\alu^{*}),& \!\!\!\!\!\!\!\!\!
\vrl&\!\!=&\!\!-\dps\jd\dps\jnl(\aqe+\aqe^{*}), \\ \;\;\;\;\;
\vrr&\!\!=&\!\!\dps\jd\dps\jnl(\aeu+\aeu^{*}),& \!\!\!\!\!\!\!\!\!
\tll&\!\!=&\!\!\dps\jc\dps\jnl\alqp^{*}, \\ \;\;\;\;\;
\trr&\!\!=&\!\!\dps\jc\dps\jnl\alqp. & & & 
\end{array}
\label{coef}
\end{equation}
We will use the following more convenient notation:
\begin{eqnarray*}
&&\ \ \ \ \ \ \ \ \ \ S\equiv S_{RR},\ \ \ T\equiv T_{RR}, \\
&&A_L\equiv V_{LL}+V_{LR},\ \ \ A_R\equiv V_{RL}+V_{RR},\\
&&B_L\equiv V_{LL}-V_{LR},\ \ \ B_R\equiv V_{RL}-V_{RR}.
\end{eqnarray*}

The differential cross section for $e^+e^-\to t\bar{t}$ as a function
of the longitudinal polarizations of electron (positron) beam ${\wj}
({\wdw})$ and of the top quark (anti-quark) spin vectors $s_+(s_-)$ 
calculated according to the lagrangian ${\cal L}={\cal L}^{S\!M}
+{\cal L}^{4F}$ is shown in appendix A. Since the electron mass is
negligible, there is no interference between scalar-tensor and vector
interactions. Therefore contributions to the cross section generated
by the scalar-tensor four-Fermi operators are of order $(\alpha_i s/
{\mit\Lambda}^2)^2$. However, the SM amplitude shall interfere with
contributions from the vector four-Fermi operators, which leads to
terms of order $\alpha_i s/{\mit\Lambda}^2$. 

\vskip 0.3cm \noindent
{\bf b. $\mib{t}$ and $\bar{\mib{t}}$ decays} \\
The following operators are found to contribute directly to decays of
top quarks:
\begin{equation}
\begin{array}{lcllcl}
{\cal O}_{qde}&\!\!=&\!\!(\bar{\ell}e)(\bar{d}q),\;\;\;&
{\cal O}_{\ell q}&\!\!=&\!\!(\bar{\ell}e)\epsilon (\bar{q}u), \\
{\cal O}_{\ell q'}&\!\!=&\!\!(\bar{\ell}u)\epsilon (\bar{q}e),\;\;\;&
{\cal O}^{(3)}_{\ell q}&\!\!=&\!\!\dps\frac12
(\bar{\ell}\gamma_{\mu}\tau^I\ell)(\bar{q}\gamma^{\mu}\tau^Iq).
\end{array}
\end{equation}
We will parameterize the corresponding lagrangian in the following
way:
\begin{eqnarray}
&&{\cal L}^{4F}=\sum_{i,j=L,R}
\Bigl[\:S_{ij}^D(\bar{\nu}_{\ell}P_i\ell)(\bar{b}P_jt) \non\\
&&\lspace
+V_{ij}^D(\bar{\nu}_{\ell}\gamma^{\mu}P_i\ell)
(\bar{b}\gamma_{\mu}P_jt)
+T_{ij}^D(\bar{\nu}_{\ell}\frac{\sigma^{\mu\nu}}{\sqrt{2}}P_i\ell)
(\bar{b}\frac{\sigma_{\mu\nu}}{\sqrt{2}}P_jt)+{\rm h.c.}\:\Bigr].
\label{dec_lag2}
\end{eqnarray}
The coefficients satisfy the constraints:
\begin{eqnarray*}
&&S^D_{LL}=S^D_{LR}=0,\ \ \ V^D_{RR}=V^D_{LR}=V^D_{RL}=0, \\
&&\ \ \ \ \ \ \ \ \ \ \ T^D_{LL}=T^D_{LR}=T^D_{RL}=0.
\end{eqnarray*}
For non-zero coefficients we get
\begin{equation}
\begin{array}{lcllcl}
\srld&\!\!=&\!\!\dps\jnl\aqde,\;\;\;&
\srrd&\!\!=&\!\!\dps\jnl(\alq-\jd\alqp), \\
\vlld&\!\!=&\!\!\dps\jnl(\alqt+\alqtc),\;\;\;&
\trrd&\!\!=&\!\!-\dps\jc\dps\jnl\alqp.
\end{array}
\end{equation}
We adopt for the notation:
$$
S^D\equiv S^D_{RR},\ \ \ V^D\equiv V^D_{LL},\ \ \ T^D\equiv T^D_{RR}.
$$

The differential decay rate for an unpolarized top quark including
both the SM and four-Fermi effective operators is given in appendix
B. In its calculations the narrow-width approximation mentioned in
the next section has been adopted. Therefore non-zero contributions
to the decay amplitude from the SM are concentrated around $(p_{\ell}
+p_\nu)^2\simeq M_W^2$ in the phase space. This means that we can
ignore interference between the SM and four-Fermi operators in the
decay. Corrections to differential decay rate are thereby of order
$(\alpha_i \mt \mw / {\mit\Lambda}^2)^2$. 

\sec{Energy spectrum of secondary leptons}

We will treat all the fermions except the top quark as massless and
adopt the technique developed by Kawasaki, Shirafuji and Tsai
\cite{formalism}. This is a useful method to calculate distributions
of final particles appearing in a production process of on-shell
particles and their subsequent decays. The technique is applicable
when the narrow-width approximation
$$
\left|\,{1\over{p^2-m^2+im{\mit\Gamma}}}\,\right|^2
\simeq{\pi\over{m{\mit\Gamma}}}\delta(p^2 -m^2)
$$
can be adopted for the decaying intermediate particles. In fact, this
is very well satisfied for both $t$ and $W$ since ${\mit\Gamma}_t
\simeq$ 175 MeV $(\mt/\mw)^3\ll\mt$ and ${\mit\Gamma}_W=2.07\pm
0.06$ GeV \cite{PDG} $\ll M_W$.

Adopting this method, one can derive the following formula for the
inclusive distribution of the single-lepton $\ell^+/\ell^-$ in the
reaction $\eett \to \ell^\pm X$ :
\begin{eqnarray}
&&\frac{1}{B_{\ell}\sigma(e^+e^-\to t\bar{t})}
\frac{d\sigma}{dx}(e^+e^-\to \ell^\pm X) \non \\
&&\ \ \ \
=\alpha_0\,[\:f(x)+(\eta^{(*)}\mp\xi^{(*)})g(x)\:]\tx \non \\
&&\lspace + \sum_{i=1}^3\alpha^{4F}_i[\:f^{4F}_i(x)+
(\eta^{(*)}\mp\xi^{(*)})g^{4F}_i(x)\:],
\label{exact}
\end{eqnarray}
where $B_{\ell}$ is the leptonic branching ratio of $t$, $r$ and
$\alpha_i^{4F}$ are defined in appendix B, $f(x)$ and $g(x)$
(Arens-Sehgal functions \cite{AS}) are recapitulated in appendix C,
the functions $f^{4F}_i(x)$ and $g^{4F}_i(x)$ $(i=1\sim 3)$ are
presented in appendix D, $x$ is the rescaled energy of the final
lepton introduced in \cite{AS}
$$
x\equiv
\frac{2 E_\ell}{\mt}\left(\frac{1-\beta}{1+\beta}\right)^{1/2}
$$
with $E_\ell$ being the energy of $\ell$ in $\epem$ c.m. frame and
$\beta\equiv\sqrt{1-4\mts/s}$, and
\beq
\eta^{(*)}\equiv\frac{4\dvak}{\mianeta},
\label{eta}
\eeq
\beq
\xi^{(*)}\equiv\frac{-2\,\zeta_2\,(3\s+4\tp)}{\mianeta},
\label{ksi}
\eeq
with $D_{V,A,V\!A}^{(*)}$ defined in appendix A and
$$
\zeta_1\equiv\fjed(1-{\wj}{\wdw}),\ \ 
\zeta_2\equiv\fjed({\wj}-{\wdw}).
$$

Below the SM-threshold $x_{th}=r(1-\beta)/(1+\beta)$ one can observe
only new-physics contributions. Therefore any non-zero signal
measured in this region must come from non-standard effects, however
it may be difficult to perform measurements for $x<x_{th}$(=0.035 for
$\sqrt{s}=500$ GeV).

\sec{Optimal-observable procedure}

Let us briefly summarize the optimal-observable procedure introduced
in ref.\cite{optobs}. Suppose we have a cross section:
$$
\frac{d\sigma}{d\phi}=\sum_ic_if_i(\phi)
$$
where $f_i(\phi)$ are known functions of the final-state phase space
$\phi$ and $c_i$ are model-dependent coefficients. These coefficients
can be extracted by using appropriate weighting functions $w_i(\phi)$
such that $\int w_i(\phi)(d\sigma/d\phi)d\phi=c_i$. There is a choice
of $w_i(\phi)$ which minimizes the resultant statistical error. Such
functions are given by
$$
w_i(\phi)=
\sum_j\frac{X_{ij}f_j(\phi)}{d\sigma(\phi)/d\phi}
$$
with $X=M^{-1}$, where
\beq
M_{ij}\equiv
\int\frac{f_i(\phi)f_j(\phi)}{d\sigma(\phi)/d\phi}d\phi.
\label{integrals}
\eeq
With these weighting functions, the statistical uncertainty of $c_i$
is estimated to be
$$
{\mit\Delta} c_i=\sqrt{X_{ii}\sigma_T/N},
$$
where $\sigma_T\equiv\int(d\sigma /d\phi)d\phi$ and $N=L_{e\!f\!f}
\sigma_T$ is the total number of events with $L_{e\!f\!f}$ being the
integrated luminosity times the detection efficiency.

Preserving only the leading terms (up to $1/{\mit\Lambda}^4$) in the
scale of new physics, one can rewrite the formula for the energy
spectrum of a single lepton in a suitable form for application of the
above optimal procedure:
\begin{equation}
\frac{1}{B_{\ell}\sigma(e^+e^-\to t\bar{t})}
\frac{d\sigma}{dx}(e^+e^-\to \ell^\pm X)=\sum_{i=1}^5c^\pm_ih_i(x)
\end{equation}
with
$$
c^\pm_1=1,\ \ 
c^\pm_2=\bar{\alpha}^{4F}_1,\ \ 
c^\pm_3=\bar{\alpha}^{4F}_2,\ \ 
c^\pm_4=\bar{\alpha}^{4F}_3,\ \
c^\pm_5={\mit\Delta}\eta\mp\bar{\xi}
$$
and
\ber
&&h_1(x)=[\:f(x)+\eta^{(*)}_{S\!M}g(x)\:]\,\theta(x-x_{th}), \\
&&h_2(x)=f^{4F}_1(x)-f(x)\,\theta(x-x_{th}) \\
&&\lspace+\eta^{(*)}_{S\!M}
\Bigl[\:g^{4F}_1(x)-g(x)\,\theta(x-x_{th})\:\Bigr],\\
&&h_3(x)=f^{4F}_2(x)-\frac13 f(x)\,\theta(x-x_{th}) \\
&&\lspace
+\eta^{(*)}_{S\!M}\Bigl[\:g^{4F}_2(x)-\frac13 g(x)\,\theta(x-x_{th})
\:\Bigr], \\
&&h_4(x)=f^{4F}_3(x)-\frac16 f(x)\,\theta(x-x_{th}) \\
&&\lspace
+\eta^{(*)}_{S\!M}\Bigl[\:g^{4F}_3(x)-\frac16 g(x)\,\theta(x-x_{th})
\:\Bigr], \\
&&h_5(x)=g(x)\,\theta(x-x_{th}),
\enr
where $\bar{\alpha}^{4F}_i(i=1\sim 3)$, $\bar{\xi}$ and $\bar{\eta}$
(used in ${\mit\Delta}\eta\equiv\bar{\eta}-\eta^{(*)}_{S\!M}$) are
the leading terms in power-series expansion (up to
$1/{{\mit\Lambda}^4}$) of $\alpha^{4F}_i\: (i=1\sim 3)$, $\xi^{(*)}$
and $\eta^{(*)}$ respectively, and $\eta^{(*)}_{S\!M}$ is the value
of $\eta^{(*)}$ in the SM.\footnote{$\eta^{(*)}_{S\!M}$ reduces to
    $\eta$ used in \cite{GH_npb,GH_plb,BGH} when $P_{e^+}=P_{e^-}=0$.
        }\ 
Notice that $c_{2,3,4}$ are of order $(\alpha_i \mt \mw /
{\mit\Lambda}^2)^2$, but $c_5$ is of order $\alpha_i s/
{\mit\Lambda}^2$ because it contains the interference part between
the SM and four-Fermi vector operators in the production $e^+e^-\to
t\bar{t}$. $h_i(x)(i=1\sim 4)$ depend on the polarization of the
initial electron and positron beams ${\wj}$ and ${\wdw}$ (through
$\eta^{(*)}_{S\!M}$).

Here we will consider both unpolarized and polarized beams, and the
polarization will be adopted to maximize non-standard effects.
{\it For illustration}, we will consider
three sets of the coefficients $\alpha_i$:
\ben
\item $\alqj=\alqt=\aeu=\alu=\aqe=\alq=\alqp=\aqde=1$,
\item $\alqj=\alqt=\aeu=\alq=\alqp=\aqde=1,\;\aqe=0,\;\alu=-1$,
\item $\alqj=\alqt=\aeu=\alu=\aqe=\alqp=\aqde=1,\;\alq=-1$ .
\een
In the following the results are given at $\sqrt{s}=500,\;750\;
\hbox{and}\; 1000$ GeV for the SM parameters $\sin^2\theta_W=0.2315$,
$m_t=175.6$ GeV, $M_W=80.43$ GeV, ${\mit\Gamma}_W=2.07$ GeV, $M_Z=
91.1863$ GeV \cite{data97}, the integrated luminosity $L=50$
fb$^{-1}$ and the single-lepton-detection efficiency $\epsilon_{\ell}
=\sqrt{0.5}$.

Since $c_{2,3,4}$ are ${\cal O}((\alpha_i \mt\mw/{\mit\Lambda}^2)^2)$
only $c_1$ and $c_5$ (${\cal O}(\alpha_i s / {\mit\Lambda}^2)$) can
be determined experimentally. Indeed we have found, for example,
\renewcommand{\arraystretch}{0.95}
$$
\baa{cllccccc} 
c_i^+&\!\!\!\!=&\!\!\!
(&\!\!\!\!1, &\!6.12\times 10^{-8}, &\!6.30\times 10^{-6},
&\!-6.36\times 10^{-6}, &\!0.717\:) \\
{\mit\Delta}c_i&\!\!\!\!=&\!\!\!
(&\!\!\!\!0.015, &\!0.021, &\!0.036, &\!0.015, &\!0.054\:)
\eaa
$$
from $e^+e^-\to \ell^+ X$ for $P_{e^+}=P_{e^-}=0.9$, ${\mit\Lambda}=
3$ TeV, $\sqrt{s}=500$ GeV and the parameter set (1). Below in Tables
\ref{tab1}, \ref{tab2} and \ref{tab3} we present $c_5^+$ and
${\mit\Delta}c_5$ calculated for two sets of $\alpha$'s (set (1) and
(2)), unpolarized and polarized beams with $\sqrt{s}=500,\;750\;
\hbox{and}\; 1000$ GeV, respectively. There, all the operators of
dimension greater than 6 have been neglected. Therefore certain
criteria for an applicability of the perturbation expansion should be
adopted. Hereafter we will present results only if the relative
correction to the total cross section for $\eett$ does not exceed
$30\%$ and $d \sigma/d x$ is always positive. The integration region
adopted in the formula (\ref{integrals}) runs from $x=0.0$ to
$x=1.0$, however in the case of a real experiment one has to adjust
it according to the detector constraints.

\begin{table}[h]
\vspace*{-0.4cm}
\bce
\begin{tabular}{||l|l|l|c|c|c||}
\hline
&$P_{e^-}$&$P_{e^+}$&\multicolumn{3}{c||}{${\mit\Lambda}$ (TeV)}\\
\cline{4-6}
&&&3&5&7\\
\hline
(1)$\;c_5$&0  &0  &0.0607&0.0345&0.0194\\
(1)$\;{\mit\Delta} c_5$&0  &0  &0.0554&0.0510&0.0484\\
(1)$\;N_{S\!D}$&0 &0 &1.0957&0.6765&0.4008\\
\hline
(1)$\;c_5$&0.9 &$-$0.9&0.1766&0.0496&0.0233\\
(1)$\;{\mit\Delta} c_5$&0.9 &$-$0.9&0.1210&0.1162&0.1108\\
(1)$\;N_{S\!D}$&0.9&$-$0.9&1.4595&0.4268&0.2103\\
\hline
(1)$\;c_5$&0.9 &0  &0.6843&0.2047&0.0986\\
(1)$\;{\mit\Delta} c_5$&0.9 &0  &0.0692&0.0624&0.0580\\
(1)$\;N_{S\!D}$&0.9&0&9.8887&3.2804&1.700\\
\hline
(1)$\;c_5$&0.9 &0.9 &0.7169&0.2125&0.1020\\
(1)$\;{\mit\Delta} c_5$&0.9 &0.9 &0.0536&0.0466&0.0432\\
(1)$\;N_{S\!D}$&0.9&0.9&13.3750&4.5601&2.3611\\
\hline
(2)$\;c_5$&0  &0  &0.3944&0.1307&0.0651\\
(2)$\;{\mit\Delta} c_5$&0  &0  &0.0700&0.0560&0.0507\\
(2)$\;N_{S\!D}$&0 &0 &5.6343 &2.3339&1.2840\\
\hline
(2)$\;c_5$&0.9 &$-$0.9&0.5103&0.1458&0.0690\\
(2)$\;{\mit\Delta} c_5$&0.9 &$-$0.9&0.1471&0.1263&0.1155\\
(2)$\;N_{S\!D}$&0.9 &$-$0.9&3.4691&1.1796&0.5974\\
\hline
(2)$\;c_5$&0.9 &0  &$-$&0.0699&0.0329 \\
(2)$\;{\mit\Delta} c_5$&0.9 &0  &$-$&0.0653&0.0592\\
(2)$\;N_{S\!D}$&0.9 &0 &$-$&1.0704&0.5557\\
\hline
(2)$\;c_5$&0.9 &0.9 &$-$&0.0411&0.0198\\
(2)$\;{\mit\Delta} c_5$&0.9 &0.9 &$-$&0.0479&0.0436\\
(2)$\;N_{S\!D}$&0.9 &0.9 &$-$&0.8580&0.4541\\
\hline
\end{tabular}
\ece
\vspace*{-0.4cm}
\caption{$c_5^+$ and ${\mit\Delta}c_5$ calculated for
$\protect\sqrt{s}=500$ GeV for various polarizations of the electron
($P_{e^-}$) and the positron ($P_{e^+}$) beam, adopting two sets ((1)
and (2)) of the coefficients $\alpha_i$. Hereafter ``$-$'' indicates
that for the parameters chosen either the correction to 
$\sigma(\epem\to\ttbar)$ exceeds $30\;\%$ or $d \sigma/d x$ becomes
negative.}
\label{tab1}
\end{table}

First of all one shall conclude from the tables that the statistical
significance of the non-standard signal (for an observation of $c_5$)
$N_{S\!D}\equiv |c_5|/{\mit\Delta}c_5$ depends strongly both on the
choice of the coefficient set and on the adopted beam polarization; 
e.g. for $\sqrt{s}=500$ GeV, $P_{e^-}=P_{e^+}=0$ and
${\mit\Lambda}=3$~TeV we read from Table~\ref{tab1} $N_{S\!D}=$1.1
and 5.6 for the set (1) and (2) respectively. The effect is caused by
an accidental cancellation in the value of $c_5$ for the set (1). 

\begin{table}[h]
\vspace*{-0.4cm}
\bce
\begin{tabular}{||l|l|l|c|c|c||}
\hline
&$P_{e^-}$&$P_{e^+}$&\multicolumn{3}{c||}{${\mit\Lambda}$ (TeV)}\\
\cline{4-6}
&&&3&5&7\\
\hline
(1)$\;c_5$             &0  &0  &$-$   &0.0782&0.0485 \\
(1)$\;{\mit\Delta} c_5$&0  &0  &$-$   &0.0522&0.0500 \\
(1)$\;N_{S\!D}$          &0  &0  &$-$   &1.4981&0.9700 \\         
\hline
(1)$\;c_5$             &0.9&$-$0.9&$-$   &0.1555&0.0686 \\
(1)$\;{\mit\Delta} c_5$&0.9&$-$0.9&$-$   &0.1155&0.1135 \\
(1)$\;N_{S\!D}$          &0.9&$-$0.9&$-$   &1.3463&0.6044 \\
\hline
(1)$\;c_5$             &0.9&0  &$-$   &0.5186&0.2228 \\
(1)$\;{\mit\Delta} c_5$&0.9&0  &$-$   &0.0668&0.0594 \\
(1)$\;N_{S\!D}$          &0.9&0  &$-$   &7.7635&3.7508 \\
\hline
(1)$\;c_5$             &0.9&0.9&$-$   &0.5257&0.2233 \\
(1)$\;{\mit\Delta} c_5$&0.9&0.9&$-$   &0.0506&0.0432 \\
(1)$\;N_{S\!D}$          &0.9&0.9&$-$   &10.3893&5.1690\\
\hline
(2)$\;c_5$             &0  &0  &0.9862&0.3111&0.1526 \\
(2)$\;{\mit\Delta} c_5$&0  &0  &0.0913&0.0608&0.0539 \\
(2)$\;N_{S\!D}$          &0  &0  &10.8018&5.1168&2.8980 \\
\hline
(2)$\;c_5$             &0.9&$-$0.9&$-$   &0.3885&0.1727 \\
(2)$\;{\mit\Delta} c_5$&0.9&$-$0.9&$-$   &0.1325&0.1222 \\
(2)$\;N_{S\!D}$          &0.9&$-$0.9&$-$   &2.9321&1.4133\\
\hline
(2)$\;c_5$             &0.9 &0  &$-$&$-$&0.0805 \\
(2)$\;{\mit\Delta} c_5$&0.9 &0  &$-$&$-$&0.0599 \\
(2)$\;N_{S\!D}$          &0.9 &0  &$-$&$-$&1.3439 \\ 
\hline
(2)$\;c_5$             &0.9 &0.9 &$-$&$-$&0.0422 \\
(2)$\;{\mit\Delta} c_5$&0.9 &0.9 &$-$&$-$&0.0415 \\
(2)$\;N_{S\!D}$          &0.9 &0.9 &$-$&$-$&1.0169 \\
\hline
\end{tabular}
\ece
\vspace*{-0.4cm}
\caption{$c_5^+$ and ${\mit\Delta}c_5$ calculated for
$\protect\sqrt{s}=750$ GeV for various polarizations of the electron
($P_{e^-}$) and the positron ($P_{e^+}$) beam adopting two sets ((1)
and (2)) of the coefficients $\alpha_i$.}
\label{tab2}
\end{table}

Comparing different choices of beam polarizations one can observe
that (especially for the set (1)) $P_{e^-}=P_{e^+}=0.9$ is by far
{\it the most convenient scenario}~\footnote{We have examined $c_5$
    dependence on $(P_{e^-},P_{e^+})$ and found that for the set (1)
    besides small areas in the vicinity of $(\pm 1.,\mp 0.9)$ the
    choice $(0.9,0.9)$ adopted in Tables~\ref{tab1},~\ref{tab2}~and~
    \ref{tab3} is indeed optimal and provides much greater $c_5$.
    However, it turns out that for the set (2) the point $(-0.9,
    -0.9)$ generates larger $c_5$ than $(0.9,0.9)$. It illustrates
    the fact that the optimal choice of polarizations depends on the
    coefficients $\alpha_i$.}\ 
since $N_{S\!D}$ could reach even $13.4$ for $\sqrt{s}=500$~GeV and
${\mit\Lambda}=3$~TeV. In fact the dominant effects from non-standard
interactions appear below the SM threshold $x_{th}(=0.035$ for
$\sqrt{s}=500$~GeV). Therefore in order to observe $N_{S\!D}$ of the
order of 13 one has to be able to detect very soft leptons. While
restricting the integration area in eq.(\ref{integrals}) to the
region above $x=0.05$ $N_{S\!D}=13$ is being reduced to $4.3$.
However, we can still conclude that physics of the scale of
${\mit\Lambda}=3$~TeV could be detected at the $\sqrt{s}=500$~GeV
collider.

\begin{table}[h]
\vspace*{-0.4cm}
\bce
\begin{tabular}{||l|l|l|c|c|c||}
\hline
&$P_{e^-}$&$P_{e^+}$&\multicolumn{3}{c||}{${\mit\Lambda}$ (TeV)}\\
\cline{4-6}
&&&3&5&7\\
\hline
(1)$\;c_5$             &0  & 0  &$-$&0.1082&0.0811 \\
(1)$\;{\mit\Delta} c_5$&0  & 0  &$-$&0.0610&0.0609 \\
(1)$\;N_{S\!D}$          &0  & 0  &$-$&1.7738&1.3317 \\
\hline
(1)$\;c_5$             &0.9&$-$0.9&$-$&$-$   &0.1449 \\
(1)$\;{\mit\Delta} c_5$&0.9&$-$0.9&$-$&$-$   &0.1354 \\
(1)$\;N_{S\!D}$          &0.9&$-$0.9&$-$&$-$   &1.0702 \\ 
\hline
(1)$\;c_5$             &0.9& 0  &$-$&1.1612 &0.4490 \\
(1)$\;{\mit\Delta} c_5$&0.9& 0  &$-$&0.0892 &0.0785 \\
(1)$\;N_{S\!D}$          &0.9& 0  &$-$&13.0179&5.7197 \\
\hline
(1)$\;c_5$             &0.9&0.9&$-$&$-$   &$-$    \\
(1)$\;{\mit\Delta} c_5$&0.9&0.9&$-$&$-$   &$-$    \\
(1)$\;N_{S\!D}$          &0.9&0.9&$-$&$-$   &$-$    \\
\hline
(2)$\;c_5$             &0  & 0  &$-$&0.5821&0.2797 \\
(2)$\;{\mit\Delta} c_5$&0  & 0  &$-$&0.0761&0.0690 \\
(2)$\;N_{S\!D}$          &0  & 0  &$-$&7.6491&4.0536 \\
\hline
(2)$\;c_5$             &0.9&$-$0.9&$-$&0.8269&0.3434 \\
(2)$\;{\mit\Delta} c_5$&0.9&$-$0.9&$-$&0.1324&0.1517 \\
(2)$\;N_{S\!D}$          &0.9&$-$0.9&$-$&6.2455&2.2637 \\    
\hline
(2)$\;c_5$             &0.9& 0  &$-$&$-$   &$-$    \\
(2)$\;{\mit\Delta} c_5$&0.9& 0  &$-$&$-$   &$-$    \\
(2)$\;N_{S\!D}$          &0.9& 0  &$-$&$-$   &$-$    \\
\hline
(2)$\;c_5$             &0.9& 0.9&$-$&$-$&$-$ \\
(2)$\;{\mit\Delta} c_5$&0.9& 0.9&$-$&$-$&$-$ \\
(2)$\;N_{S\!D}$          &0.9& 0.9&$-$&$-$&$-$ \\
\hline
\end{tabular}
\ece
\vspace*{-0.4cm}
\caption{$c_5^+$ and ${\mit\Delta}c_5$ calculated for
$\protect\sqrt{s}=1000$ GeV for various polarizations of the electron
($P_{e^-}$) and the positron ($P_{e^+}$) beam adopting two sets ((1)
and (2)) of the coefficients $\alpha_i$.}
\label{tab3}
\end{table}

We have checked that adopting $4\sigma$ as a discovery signal we can
conclude that if the set (1) was chosen by Nature one would be able
to detect deviations from the SM even if the scale of non-standard
interactions was approximately $5$ times larger than $\sqrt{s}$,
adopting $P_{e^-}=P_{e^+}=0.9$ and restricting the integration region
to $0.05<x<1.0$! It should be emphasized that such a large $N_{S\!D}$
could be reached keeping the non-standard correction to
$\sigma(\epem\to\ttbar)$ below $30\;\%$! One should also notice that
even for unpolarized positron beam, for the set (1), $\sqrt{s}=500$~
GeV, $P_{e^-}=0.9$ and for restricted integration region one can
expect $N_{S\!D}=3.3 \; \hbox{and} \; 1.1$ for ${\mit\Lambda}=3 \;
\hbox{and} \; 5 $~TeV, respectively.

For polarized-initial-lepton beams a useful measure of contributions
from the scalar-tensor four-Fermi operators in the production could
be the energy-spectrum asymmetry $a(x)$ introduced in \cite{AS,CKP},
which is given by
\begin{equation}
a(x)\equiv\frac{d\sigma^-/dx-d\sigma^+/dx}{d\sigma^-/dx+d\sigma^+/dx}
=\xi^{(*)}\frac{g(x)}{f(x)+\eta^{(*)} g(x)}\ \ \ ({\rm for}\ x\geq
x_{th})
\end{equation}
in our approximation. 
\begin{figure}[h]
\vspace*{-4.5cm}
\postscript{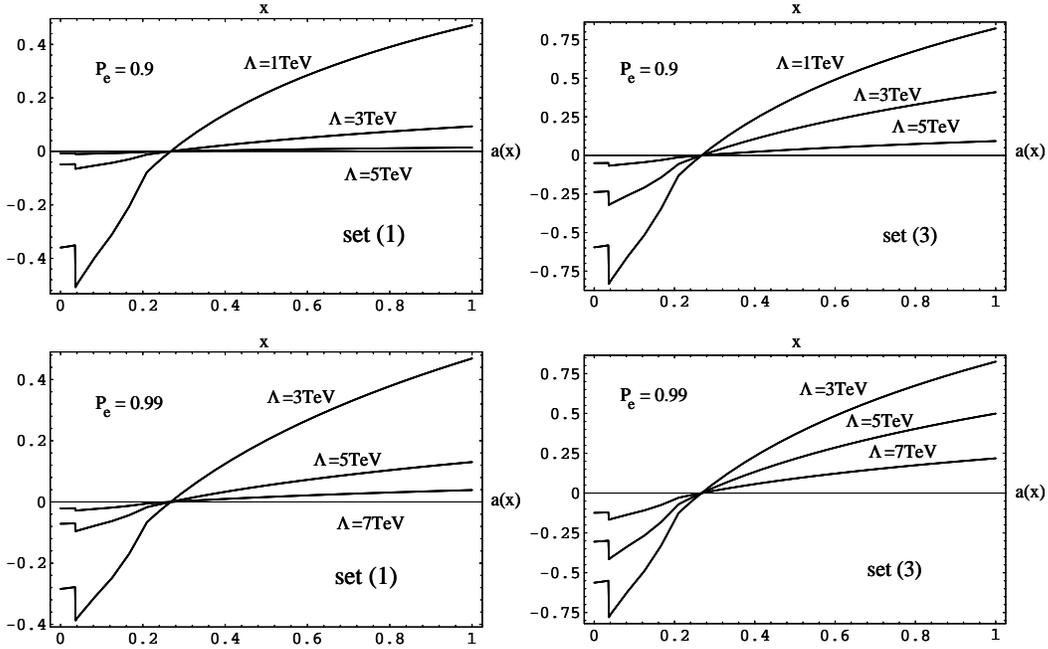}{1.0}
\vspace*{-5cm}
\caption{The asymmetry $a(x)$ for initial polarization 
$P_e(=P_{e^-}=-P_{e^+})=0.9 \; \hbox{and} \; 0.99$, 
$\protect\sqrt{s}=500$ GeV, ${\mit\Lambda}=1\sim 7$ TeV and for the
coefficient sets (1) and (3). The step-function-like change in the
curves at $x=0.035$ is due to the SM-threshold.} 
\label{asymmetry}
\end{figure}
Indeed the asymmetry seems to be a good measure of $\xi^{(*)}$ which
receives contributions only from the scalar-tensor operators. It
should be noticed, however, that the value of $\xi^{(*)}$ depends
very strongly on initial-lepton-beam polarizations, effectively it is
non-vanishing only in the vicinity of $P_{e^\pm}=1$; at least one
beam must be polarized. Figure~\ref{asymmetry} shows
$a(x)\,$\footnote{Calculated according to the general form of
    $d\sigma/dx$ given by the equation (\ref{exact}).}\ for various
values of ${\mit\Lambda}$, $P_e(=\wj=-\wdw)=0.9$ and 0.99, and two
coefficient sets (1) and (3). Here the coefficient set (3) has been
adopted to avoid an accidental cancellation between $\alpha_{lq}$ and
$\alpha_{lq'}$ in the value of $S_{LL}$ (see eq.(\ref{coef})). In
fact it is seen from the figure that the asymmetry for the set (3)
gains an extra factor of about 2 in comparison with the set (1). An
increase of $P_e$ enhances the relative strength of the new-physics
effects (from scalar- and tensor-operators), because the opposite
polarization of initial $e^{\pm}$ beams reduces the SM (or more
generally vector-operator) contribution. Thus it causes an
intensification of $a(x)$ dependence on the new-physics energy scale
${\mit\Lambda}$, as seen from the figure. Therefore large $P_e$
allows to penetrate higher energy scales.

\begin{figure}[h]
\vspace*{-4.5cm}
\postscript{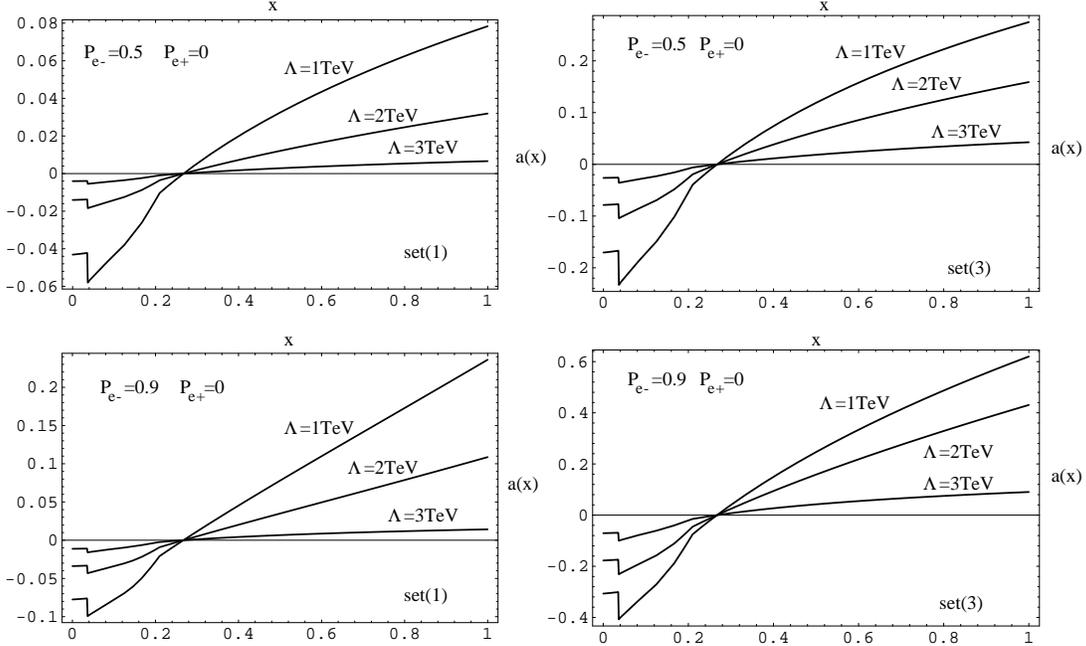}{1.0}
\vspace*{-5cm}
\caption{The asymmetry $a(x)$ for unpolarized positron and polarized
electron beam ($P_{e^-}=0.5 \; \hbox{and} \; 0.9$), $\protect\sqrt{s}
=500$ GeV, ${\mit\Lambda}=1\sim 3$ TeV and for the coefficient sets
(1) and (3).} 
\label{asymmetry_new}
\end{figure}

One should, however, keep also in mind that increasing {\it opposite
polarization of both beams} we suppress the (SM-like) vector-operator
contributions and therefore the total number of events is strongly
reduced, see Tab.\ref{crossec}, so the measurement of the asymmetry
will be a challenging task for experimentalists. Therefore it is
instructive to consider unpolarized positron beam. Besides, in
practice it appears to be much more difficult to achieve positron
polarization, so below we also present results for unpolarized
positron beams.

\begin{table}[h]
\vspace*{-0.4cm}
\bce 
\begin{tabular}{||l|c|c|c|c||}
\hline
$P_e$&\multicolumn{4}{c||}{${\mit\Lambda}$ (TeV)}\\
\cline{2-5}
&3&5&7&SM\\
\hline
0   &0.68&0.61&0.59&0.58 \\
0.5 &0.51&0.46&0.45&0.44 \\
0.9 &0.14&0.12&0.11&0.11 \\
0.99&0.03&0.01&0.01&0.01 \\
\hline
\end{tabular}\\
\ece
\vspace*{-0.5cm}
\caption{The total cross section $\sigma(e^+e^-\to t\bar{t})$ in pb
with $\protect\sqrt{s}=500$ GeV, for ${\mit\Lambda}=3,\:5,\:7$ TeV 
with the coefficient set (1) and the SM, for polarization 
$P_e(=P_{e^-}=-P_{e^+})=0.0,\;0.5,\;0.9,\;0.99$. 
Here we used $\alpha(s) (\simeq 1/126)$ instead of $\alpha(0)$.} 
\label{crossec}
\end{table}

It is seen from the plots in Figs.\ref{asymmetry},
\ref{asymmetry_new} that the typical size of the asymmetry for
unpolarized positrons is smaller than the one for opposite electron
and positron polarization, therefore sensitivity of the asymmetry to
non-standard physics embedded in the coefficients $S$ and $T$ is
being reduced. The reason is that for $P_e^+=0$ the parameter
$\xi^{(*)}$ defined by eq.(\ref{ksi}) is suppressed by non-zero SM
contributions. However, one can observe that for the set (3),
$P_{e^-}=0.9$, $\sqrt{s}=500$~GeV and ${\mit\Lambda}=1$~TeV the
asymmetry could be still large, of the order of $50\%$. One can
conclude that in order to penetrate physics up to ${\mit\Lambda}=2$~
TeV at $\sqrt{s}=500$~GeV electron polarization greater than $P_{e^-}
=0.9$ would be needed.

\sec{Summary}

Next-generation linear colliders of $e^+ e^-$, NLC, will provide the
cleanest environment for studying top-quark interactions. There, we
shall be able to perform detailed tests of top-quark couplings and
either confirm the SM simple generation-repetition pattern or
discover some non-standard interactions. In this paper, we focused on
the four-Fermi-type new interactions, and studied their possible
effects in $e^+e^-\to t\bar{t}\to \ell^\pm \cdots$ for arbitrary
longitudinal beam polarizations. Then, the recently proposed
optimal-observables technique \cite{optobs} has been adopted to
determine non-standard couplings through single-leptonic-spectrum
measurements.

There are scalar-, vector- and tensor-type four-Fermi interactions
contributing to our process. Since the first and last ones do not
interfere with the standard contribution, their effects were found
too small to be detected directly in the secondary-lepton-energy
spectrum, though the details depend on the size of the new-physics
scale ${\mit\Lambda}$. On the other hand, the vector interactions can
interfere with the SM contributions, so there seems to be a chance to
detect their effects through the optimal observables if
${\mit\Lambda}$ is not too high; e.g. ${\mit\Lambda}\lesim 3 $~TeV 
may provide  $4\sigma$ effects at $\sqrt{s}=500$~GeV.

In order to detect a signal of the scalar- and tensor-interactions,
we considered the lepton-energy asymmetry $a(x)$. We conclude that
the asymmetry might be useful when we achieve highly polarized
$e^\pm$ beams. Indeed, we found that at $\sqrt{s}=500$~GeV $a(x)$
becomes of the order of $25\,\%$ even for ${\mit\Lambda}=3$~TeV when
both beams are polarized simultaneously to $P_e(=P_{e^+}=P_{e^-})=
0.9$. High polarization of positron beam is hard to realize, however
we found that the use of polarized $e^-$ beam is still effective
even when $P_{e^+}=0$. For example, the size of $a(x)$ could reach
$50\%$ for ${\mit\Lambda}=1$~TeV for $\wj=0.9$.

\vskip 0.6cm
\centerline{ACKNOWLEDGMENTS}

\vspace*{0.3cm}
This work is supported in part by the State Committee for Scientific
Research (Poland) under grant 2 P03B 180 09 and by Maria
Sk\l odowska-Curie Joint Fund II (Poland-USA) under grant
MEN/NSF-96-252.

\vspace*{0.7cm}
\def\sec#1{\addtocounter{section}{1}\section*{\hspace*{-0.72cm}
\normalsize\bf Appendix \Alph{section}.$\;$#1}\vspace*{-0.3cm}}
\setcounter{section}{0}
\noindent
\sec{Differential cross section for $\mib{e}^+\mib{e}^-\to \mib{t}
\bar{\mib{t}}$}

The differential cross section for $e^+e^-\to t\bar{t}$ as a function
of $P\equiv p_{e^-}+p_{e^+}$, $l\equiv p_{e^-}-p_{e^+}$, $q\equiv
p_t-p_{\bar{t}}$, the longitudinal polarization ${\wj}({\wdw})$ of
the initial electron (positron) beam and spin 4-vectors $s_+(s_-)$ of
$t(\bar{t})$ taking into account corrections from four-Fermi
operators (\ref{dec_lag}) is given by the following formula:\\
(1) Scalar-Tensor operators : 
\begin{eqnarray}
&&\frac{d\sigma}{d\Omega_t}^{\!\!ST}
\!\!\!\!({\wj},{\wdw},s_+,s_-) \non \\
&&=\frac{3\beta}{512\pi^2 s}\Bigl[\:\:(1-{\wj}{\wdw})
[\:\s s\{s-2m_t^2(1-s_+s_-)\} \non\\
&&\ \ \ \ \ \ \ \lspace
+4\,\tp\{2m_t^2 s(1-s_+s_-)+(\invlq)^2+4m_t^2(\invppll)\} \non\\
&&\ \ \ \ \ \ \ \lspace
+4\,\rest\{\invlq\:s +2m_t^2(\invlppl)\} \non\\
&&\ \ \ \ \ \ \ \lspace
+8\,\imst m_t^2\epsilon(s_+,s_-,P,l)\:] \non\\
&&\ \ \ \ \ \ \
-2({\wj}-{\wdw})m_t[\:\s s(\invP)+4\,\tp\invlq(\invl)\non\\
&&\ \ \ \ \ \ \ \lspace
+2\,\rest\{s(\invl)+\invlq(\invP)\} \non\\
&&\ \ \ \ \ \ \ \lspace
+2\,\imst\{\epsilon(s_+,P,q,l)+\epsilon(s_-,P,q,l)\}\:]\:\:\Bigr].
\end{eqnarray}
(2) Standard Model plus Vector operators :
\begin{eqnarray}
&&\frac{d\sigma}{d\Omega_t}^{\!\!S\!M+V}
\!\!\!\!\!\!\!\!\!\!({\wj},{\wdw},s_+,s_-)
\non\\
&&=\frac{3\beta\alpha^2}{16 s^3}\:\Bigl[
\:\:D_V^{(*)}\:
[\:\{ 4m_t^2s+(lq)^2 \}(1-s_{+}s_{-})+s^2(1+s_{+}s_{-}) \non\\
&&\ \ \ \lspace +2s(ls_{+}\;ls_{-}-Ps_{+}\;Ps_{-})
           +\:2\,lq(ls_{+}\;Ps_{-}-ls_{-}\;Ps_{+})\:]   \non\\
&&\ \ \
+\:D_A^{(*)}\:[\:(lq)^2(1+s_{+}s_{-})-(4m_t^2s-s^2)(1-s_{+}s_{-})
\non\\
&&\ \ \ \lspace -2(s-4m_t^2)(ls_{+}\;ls_{-}-Ps_{+}\;Ps_{-})
           -\:2\,lq(ls_{+}\;Ps_{-}-ls_{-}\;Ps_{+})\:]   \non\\
&&\ \ \ 
-4\:{\rm Re}(D_{V\!\!A}^{(*)})\:m_t\,
[\:s(\psp-\psm)+lq(\lsp+\lsm)\:]                        \non\\
&&\ \ \ 
+2\:{\rm Im}(D_{V\!\!A}^{(*)})\:[\:lq\,\eps(\splus,\smin,q,l)
+\lsm\eps(\splus,P,q,l)+\lsp\eps(\smin,P,q,l)\:]        \non\\
&&\ \ \
+4\:E_V^{(*)}\:\m s(\lsp+\lsm)+4\:E_A^{(*)}\:\m\,lq(\psp-\psm) \non\\
&&\ \ \ 
+4\:{\rm Re}(E_{V\!\!A}^{(*)})\:
[\:2\mq(\lsp\;\psm-\lsm\;\psp)-lq\:s\:]                 \non\\
&&\ \ \ +4\:{\rm Im}(E_{V\!\!A}^{(*)})\:\m[\:\eps(\splus,P,q,l)
+\eps(\smin,P,q,l)\:]\:\:\Bigr],
\end{eqnarray}
where
\begin{eqnarray*}
&&\dvk=\jww\Bigl[\:\fdv\:\Bigr] \\
&&\lspace-2\,\ww \fev, \\
&&\dak=\jww\Bigl[\:\fda\:\Bigr]  \\
&&\lspace-2\,\ww \fea, \\
&&\dvak=\jww\Bigl[\:\fdva\:\Bigr] \\
&&\lspace-\ww \Bigl[\:\feva\:\Bigr], \\
&&\evk=2\,\jww \fev \\
&&\lspace-\ww \Bigl[\:\fdv\:\Bigr], \\
&&\eak=2\,\jww \fea \\
&&\lspace-\ww \Bigl[\:\fda\:\Bigr], \\
&&\evak=\jww\Bigl[\:\feva\:\Bigr]\\
&&\lspace-\ww\Bigl[\:\fdva\:\Bigr],
\end{eqnarray*}
with
$$
d\equiv\frac{s}{s-M_Z^2}\frac{1}{16\sktw\cktw},
$$
$$
v_f\equiv 2I^f_3-4e_f\sktw,\ \ a_f\equiv 2I^f_3,
$$
$I^f_3=\pm 1/2$ for up or down particles, and $e_f$ is an electric
charge in units of the electric charge of the proton. The symbol
$\epsilon(a,b,c,d)$ means $\epsilon_{\mu\nu\rho\sigma}a^{\mu}b^{\nu}
c^{\rho}d^{\sigma}$ with $\epsilon_{0123}=+1$. The longitudinal
polarizations of electrons and positrons are by definition:
$$
{\wj}=\frac{\njp-\njm}{\njp+\njm},\ \ \
{\wdw}=-\frac{\ndp-\ndm}{\ndp+\ndm}
$$
with
\begin{center}
$\bullet~\njp$~number~of~electrons~with\ helicity~+ \\
$\bullet~\njm$~number~of~electrons~with\ helicity~$-$ \\
$\bullet~\ndp$~number~of~positrons~with\ helicity~+ \\
$\bullet~\ndm$~number~of~positrons~with\ helicity~$-$
\end{center}

\sec{Differential decay rate for an unpolarized top quark}

The differential decay rates for an unpolarized $t$ and $\bar{t}$
quark including the Standard Model and four-Fermi operators
(\ref{dec_lag2}) are both given by:
\begin{eqnarray}
&&\frac{1}{{\mit\Gamma}_t}\frac{d^2{\mit\Gamma}_{\ell}}{dx d\omega}
(\stackrel{\scriptscriptstyle(-)}t\!(p_t)\to \ell^\pm (p_{\ell})X)
\non \\
&&\ \ \ 
=\frac{1+\beta}{\beta}B_{\ell}
\Bigl[\:\frac{3}{W}\alpha_0\omega\tom\tx
+\sum_{i=1}^{3}\alpha^{4F}_i\omega^{i-1}\:\Bigr], \label{d-width}
\end{eqnarray}
where $\omega \equiv (p_t -p_{\ell})^2/m_t^2$, ${\mit\Gamma}_t$ is
the total width of $t$,
\begin{eqnarray*}
&&\alpha_0=\frac{G}{\mianalfa},\\
&&\alpha^{4F}_1=2\frac{\sd+\sdrl+4\td+4\st}{\mianalfa},\\
&&\alpha^{4F}_2=2\frac{\sd +\sdrl +24\vd +52\td -20\st}{\mianalfa},\\
&&\alpha^{4F}_3=-4\frac{\sd +\sdrl +12\vd +28\td -8\st}{\mianalfa},
\end{eqnarray*}
and
\[
G\equiv \fdwa,\ \ \ W\equiv(1-r)^2(1+2r),\ \ \ r\equiv(M_W/m_t)^2.
\]
Note that $\alpha_0$ and $\alpha^{4F}_i$ satisfy
\begin{equation}
\alpha_0+\alpha^{4F}_1+\frac{1}{3}\alpha^{4F}_2+
\frac16\alpha^{4F}_3=1.
\label{identity}
\end{equation}
As is seen from $\alpha_0$ and $\alpha^{4F}_i$, the first term in
eq.(\ref{d-width}) (with two $\theta$-functions) is the SM
contribution and the second term is from the four-Fermi operators.
Since we used the narrow-width approximation in the SM part, the
ranges of $x$ and $\omega$ there are different from those in the
second term. The two $\theta$-functions express this difference. See
appendices C and D for more details.

\sec{Functions $\mib{f}(\!\!(\mib{x})\!\!)$ and
$\mib{g}(\!\!(\mib{x})\!\!)$}

The functions $f(x)$ and $g(x)$ are defined as
\begin{equation}
f(x)\equiv\frac{3}{W}\frac{1+\beta}{\beta}\int d\omega\,\omega,
\end{equation}
\begin{equation}
g(x)\equiv\frac{3}{W}\frac{1+\beta}{\beta}\int d\omega\,\omega
\Bigl[\:1-\frac{x(1+\beta)}{1-\omega}\:\Bigr].
\end{equation}
The variable $\omega$ is constrained by the inequalities
$$
0\leq\omega\leq 1-r \ \ \ {\rm and}\ \ \
1-x\frac{1+\beta}{1-\beta}\leq\omega\leq 1-x
$$
while the reduced energy is bounded by
$$
r\frac{1-\beta}{1+\beta}\leq x\leq 1.
$$

Carrying out the integration yields
\def\Ca{{3\over{W}}{{1+\beta}\over{2\beta}}}
\def\Cb{{3\over{W}}{{(1+\beta)^2}\over\beta}}
\begin{eqnarray*} \hspace*{2cm}
&&f(x)=\Ca\,\Bigl[\: r(r-2)+2x{{1+\beta}\over{1-\beta}}
-x^2 \Bigl({{1+\beta}\over{1-\beta}}\Bigr)^2\: \Bigr], \\
&&\hspace*{7.84cm}({\rm for\ the\ interval}\ I_1,\ I_4) \\
&&\phantom{f(x)}=\Ca \, (1-r)^2,
  \hspace*{3.22cm}({\rm for\ the\ interval}\ I_2) \\
&&\phantom{f(x)}=\Ca \, (1-x)^2,
  \hspace*{3.18cm}({\rm for\ the\ interval}\ I_3,\ I_6) \\
&&\phantom{f(x)}=\frac6W\frac{1+\beta}{(1-\beta)^2}\,
x(1-\beta-x),
  \hspace*{1.85cm}({\rm for\ the\ interval}\ I_5)
\end{eqnarray*}
\begin{eqnarray*} \hspace*{1.3cm}
&&g(x)=\Cb\, \Bigl[\: -rx +x^2 {{1+\beta}\over{1-\beta}}
-x\ln {{x(1+\beta)}\over{r(1-\beta)}} \\
&&\phantom{g(x)}\ \ \ \ \ \ \ \
+{1\over{2(1+\beta)}}\Bigl\{\: r(r-2)+2x{{1+\beta}\over{1-\beta}}
-x^2 \Bigl({{1+\beta}\over{1-\beta}}\Bigr)^2\: \Bigr\}\: \Bigr], \\
&&\hspace*{8.5cm}({\rm for\ the\ interval}\ I_1,\ I_4) \\
&&\phantom{g(x)}
=\Cb\, \Bigl[\: (1-r+\ln r)x +{1\over{2(1+\beta)}}(1-r)^2 \:\Bigr],\\
&&\hspace*{8.5cm}({\rm for\ the\ interval}\ I_2) \\
&&\phantom{g(x)}
=\Cb\, \Bigl[\: (1-x+\ln x)x +{1\over{2(1+\beta)}}(1-x)^2 \:\Bigr],\\
&&\hspace*{8.5cm}({\rm for\ the\ interval}\ I_3,\ I_6) \\
&&\phantom{g(x)}
=\frac3W\frac{1+\beta}{\beta(1-\beta)^2}\,x
\Bigl[\: 2\beta(1-\beta-\beta^2 x)
-(1+\beta)(1-\beta)^2 \ln{{1+\beta}\over{1-\beta}} \:\Bigr], \\
&&\hspace*{8.5cm}({\rm for\ the\ interval}\ I_5)
\end{eqnarray*}
where $I_i(i=1\sim 6)$ are given by
\ber
&&I_1:r\jmpb\leq x\leq \jmpb,\\
&&I_2:\jmpb\leq x\leq r,\\
&&I_3:r\leq x\leq 1,
\enr
($I_{1,2,3}$ are for $r\geq \jmpb$)
\ber
&&I_4:r\jmpb\leq x\leq r,\\
&&I_5:r\leq x\leq \jmpb,\\
&&I_6:\jmpb\leq x\leq 1,
\enr
($I_{4,5,6}$ are for $r\leq \jmpb$).\\
Note that $f(x)$ and $g(x)$ satisfy
\begin{equation}
\int f(x)dx=1,\ \ \ \ \int g(x)dx=0.
\end{equation}

\sec{Functions $\mib{f}^{{\bf 4}F\!\!\!\!\!F}(\!\!(\mib{x})\!\!)$
and $\mib{g}^{{\bf 4}F\!\!\!\!\!\!\,F}(\!\!(\mib{x})\!\!)$}

The functions $f^{4F}_i(x)$ and $g^{4F}_i(x)$ (for $i=1\sim 3$)
are defined as
\begin{equation}
f^{4F}_i(x)\equiv\frac{1+\beta}{\beta}\int d\omega\,\omega^{i-1},
\end{equation}
\begin{equation}
g^{4F}_i(x)\equiv\frac{1+\beta}{\beta}\int d\omega\,\omega^{i-1}
\Bigl[\:1-\frac{x(1+\beta)}{1-\omega}\:\Bigr].
\end{equation}
The variable $\omega$ is constrained by the inequalities
$$
0\leq\omega\leq 1 \ \ \ {\rm and}\ \ \
1-x\frac{1+\beta}{1-\beta}\leq\omega\leq 1-x
$$
while the reduced energy is bounded by
$$
0\leq x\leq 1.
$$

Carrying out the integration yields
\renewcommand{\theequation}{\ for\ the\ interval\ $I^{4F}_1$\ }
\addtocounter{equation}{-1}
\begin{eqnarray}
&&f^{4F}_1(x)=\frac{2(1+\beta)}{1-\beta}\,x,\non\\
&&f^{4F}_2(x)=\frac{2(1+\beta)}{(1-\beta)^2}\,x(1-\beta-x),\non\\
&&f^{4F}_3(x)=\frac{2(1+\beta)}{3(1-\beta)^3}
\,x\bigl[\: 3(1-\beta)(1-\beta-2x) \non\\
&&\lspace\lspace +(3+\beta^2)x^2 \:\bigr],
\end{eqnarray}
\renewcommand{\theequation}{\ for\ the\ interval\ $I^{4F}_2$\ }
\addtocounter{equation}{-1}
\begin{eqnarray}
&&f^{4F}_1(x)=\frac{1+\beta}{\beta}(1-x),\non\\
&&f^{4F}_2(x)=\frac{1+\beta}{2\beta}(1-x)^2,\non\\
&&f^{4F}_3(x)=\frac{1+\beta}{3\beta}(1-x)^3,~~~~~~~~~~~
\end{eqnarray}
\renewcommand{\theequation}{\ for\ the\ interval\ $I^{4F}_1$\ }
\addtocounter{equation}{-1}
\hspace*{-1.5cm}
\begin{figure}[h]
\vspace*{-3.5cm}
\postscript{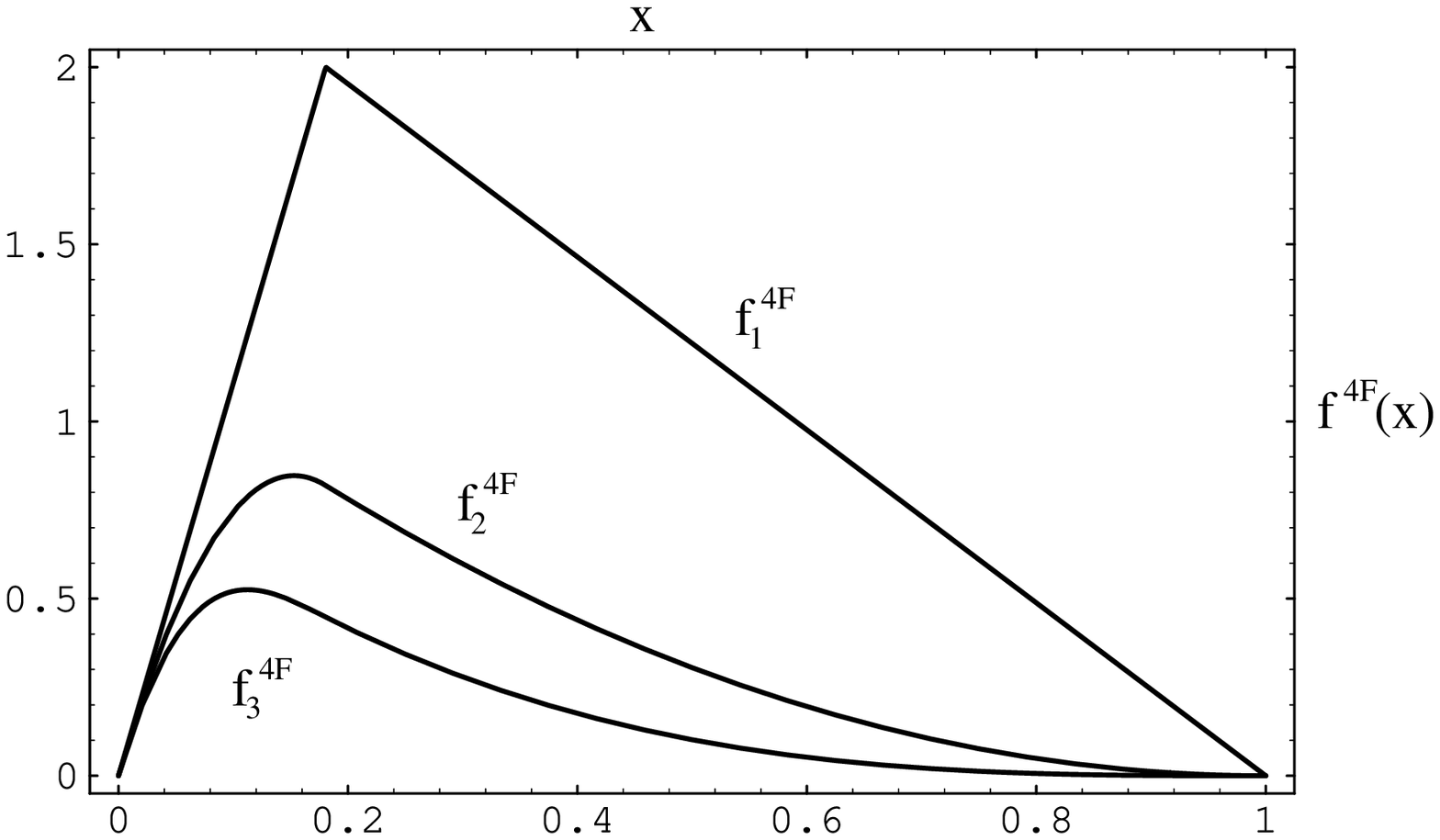}{0.90}
\vspace*{-1.5cm}
\caption{The functions $f^{4F}_i(x)$.}
\label{f4f}
\end{figure}
\begin{eqnarray}
&&g^{4F}_1(x)=\frac{1+\beta}{\beta(1-\beta)}\,x
 \Bigl[\:2\beta+(1-\beta^2)\ln\frac{1-\beta}{1+\beta}\:\Bigr],\non\\
&&g^{4F}_2(x)=\frac{1+\beta}{\beta(1-\beta)^2}\,x
 \Bigl[\:2\beta(1-\beta-\beta^2x)
 +(1+\beta)(1-\beta)^2\ln\frac{1-\beta}{1+\beta}\:\Bigr], \non\\
&&g^{4F}_3(x)=\frac{1+\beta}{3\beta(1-\beta)^3}\,x
 \Bigl[\:6\beta(1-\beta)(1-\beta-2\beta^2x)+8\beta^3x^2 \non\\
&&\lspace\lspace
 +3(1+\beta)(1-\beta)^3 \ln\frac{1-\beta}{1+\beta}\:\Bigr],
\end{eqnarray}
\renewcommand{\theequation}{\ for\ the\ interval\ $I^{4F}_2$\ }
\addtocounter{equation}{-1}
\begin{eqnarray}
&&g^{4F}_1(x)=\frac{1+\beta}{\beta}
  \bigl[\,1-x+(1+\beta)x\ln\,x\,\bigr],\non\\
&&g^{4F}_2(x)
  =\frac{1+\beta}{2\beta}\bigl[\,1+2\beta x-(1+2\beta)x^2+
  2(1+\beta)x\ln\,x\,\bigr],\non\\
&&g^{4F}_3(x)=\frac{1+\beta}{6\beta}
  \bigl[\,2+3(1+3\beta)x -6(1+2\beta)x^2+(1+3\beta)x^3\non\\
&&\lspace\lspace +6(1+\beta)x\ln\,x\,\bigr],~~
\end{eqnarray}
\renewcommand{\theequation}{\arabic{equation}}
where $I_i^{4F}\:(i=1,\:2)$ are given by
$$
I_1^{4F}:\ 0\leq x\leq \frac{1-\beta}{1+\beta},\ \ \
I_2^{4F}:\ \frac{1-\beta}{1+\beta}\leq x\leq 1.
$$
Note that $f^{4F}_i(x)$ and $g^{4F}_i(x)$ satisfy
\begin{equation}
\begin{array}{ll}
\dps\int f^{4F}_1(x)dx=1,~~~&\dps\int f^{4F}_2(x)dx=\dps\frac13,\\
\dps\int f^{4F}_3(x)dx=\dps\frac16,&\dps\int g^{4F}_i(x)dx=0,
\end{array}
\end{equation}
for $i=1\sim 3$.

\begin{figure}[h]
\vspace*{-1.5cm}
\postscript{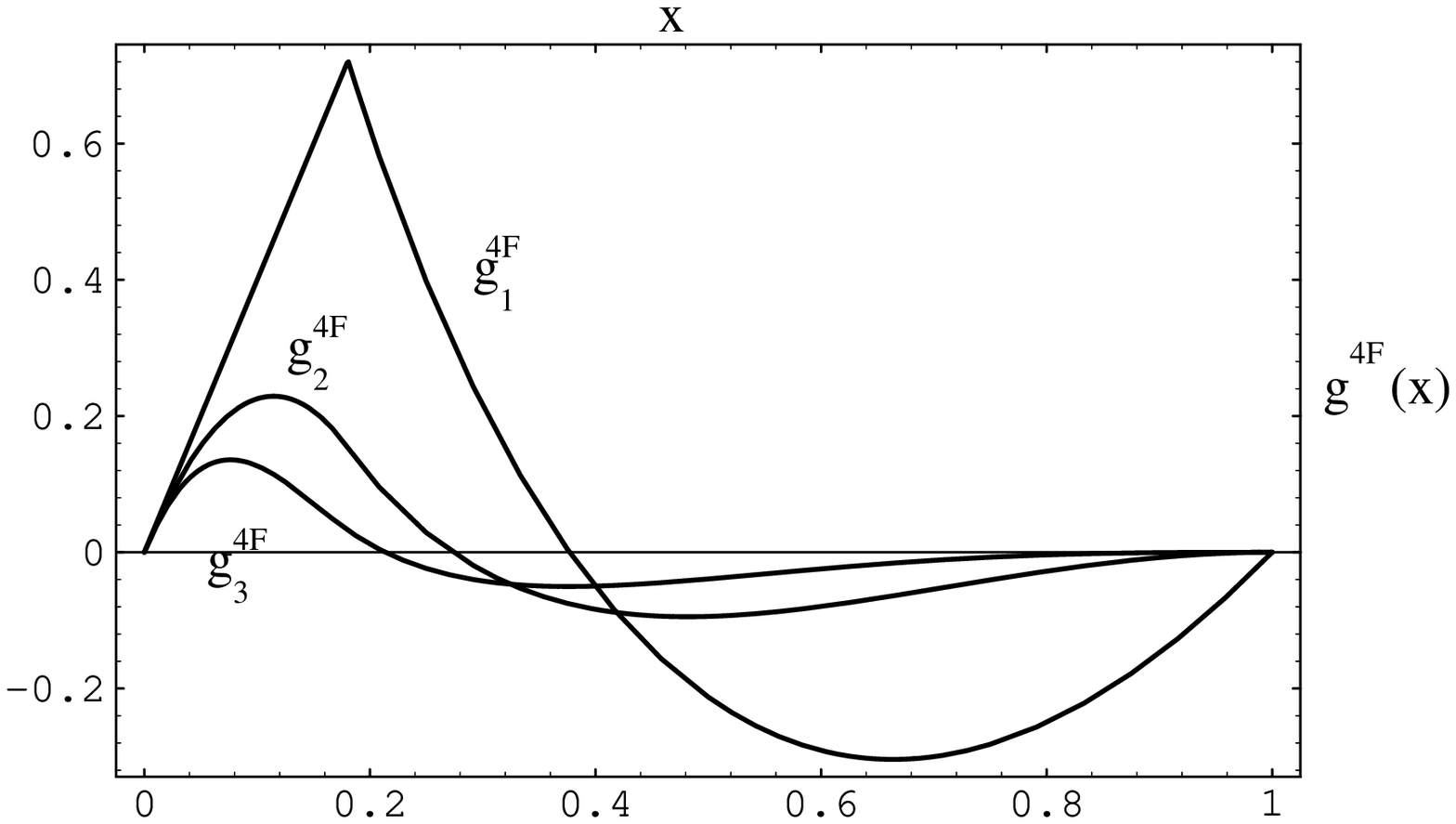}{0.90}
\vspace*{-1.5cm}
\caption{The functions $g^{4F}_i(x)$.}
\label{g4f}
\end{figure}

\newpage 

\end{document}